\documentclass[twocolumn,showpacs,preprintnumbers,amsmath,amssymb]{revtex4}

\usepackage{graphicx}
\usepackage{dcolumn}
\usepackage{bm}

\begin{document}


\title{Computing the monomer-dimer systems through matrix permanent}

\author{Yan Huo}
\author{Heng Liang}%
\author{Si-Qi Liu}%

\author{Fengshan Bai}
\email{fbai@math.tsinghua.edu.cn}
\affiliation{
Department of Mathematical Sciences, Tsinghua University, Beijing, China, 100084 
}%

\date{\today}

\begin{abstract}
The monomer-dimer model is fundamental in statistical mechanics.
However, it is $\#P$-complete in computation, even for two
dimensional problems. A formulation in matrix permanent for the
partition function of the monomer-dimer model is proposed in this
paper, by transforming the number of all matchings of a bipartite
graph into the number of perfect matchings of an extended bipartite
graph, which can be given by a matrix permanent. Sequential
importance sampling algorithm is applied to compute the permanents.
For two-dimensional lattice with periodic condition, we obtain $
0.6627\pm0.0002$, where the  value is $h_2=0.662798972834$. For
three-dimensional problem, our numerical result is $
0.7847\pm0.0014$, which agrees with the best known
bound. 
\end{abstract}

\pacs{05.50.+q, 02.10.Ox, 02.70.Uu, 02.50.-r}
\maketitle

\section{Introduction}

The monomer-dimer model is considered, in which the set of sites in
a lattice is covered by a non-overlapping arrangement of monomers
(molecules occupying one site) and dimers (molecules occupying two
sites that are neighbors in the lattice). It is fundamental in
lattice statistical mechanics \cite{HL72,KRS96}. A two dimensional
monomer-dimer model with size $m=(m_1, m_2)$ is a rectangle lattice
with $m_1\times m_2$ sites. The two dimensional monomer-dimer
systems are used to investigate the properties of adsorbed diatomic
molecules on a crystal surface \cite{Rob35}; the three-dimensional
systems occur classically in the theory of mixtures of molecules of
different sizes \cite{Gugg52} as well as the cell cluster theory of
the liquid state \cite{CBS55}. More complete description of the
history and the significance of monomer-dimer model can be found in
\cite{HL72} and the references therein.

All possible monomer-dimer coverings for a given lattice defines the
{\em configuration space} of a monomer-dimer model. { A fundamental
question for such a statistical mechanics model is to determine the
cardinal number of the  configuration space.} Practically, most of
the thermodynamic properties of  physical systems can be obtained
from the number of all possible ways that a given lattice can be
covered. Thus a considerable attention has been devoted to such a
counting problem. For a d-dimensional cubic lattice with size
$m=(m_1, m_2, \cdots , m_d)$,  this cardinal number is denoted by
$Z(m,d)$. It is proved that the following limit exists
\[h_d=\lim_{m\to\infty}\frac{\log Z(m,d)}{m_1 m_2 \cdots m_d}.\]
The limit $h_d$ is called monomer-dimer constant  \cite{Ham66}.

Even for the simplest two dimensional models, there are very few
closed form results on the monomer-dimer constant. Baxter and Gaunt
\cite{Bax68,Gau69} gives  estimates of the constants using the
asymptotic expansions. Hammersley and Menon \cite{HM70} estimate the
$h_2$ by calculating lower and upper bounds.  Numerical simulation
should play a very important role. However it has been proved that
computing the monomer-dimer constant is a $\#P$-complete problem
even for 2-dimensional problems \cite{Jer87}, which shows the
hardness of the computation. The Monte Carlo method is applied to
study the problem in \cite{KRS96,Ham66,BOS01}, which is a natural
consideration. Recently, Friedland and Peled \cite{FP05} give a
complete up-to-date theory of the computation of monomer-dimer
constant by calculating lower and upper bounds. They obtain  $h_2 =
0.66279897$, which agrees with the heuristic estimation
$e^{h_2}=1.940215351$ due to Baxter \cite{Bax68}, and $0.7653 \leq
h_3 \leq 0.7862$. Two-dimensional model with fixed dimer density is
studied intensively by Kong \cite{Kong06}. { The monomer-dimer
constant with 12 digits accuracy for two dimensional problem is
given as $h_2=0.662798972834$.}

In this paper, we propose a formulation that transforms the counting
of all matchings of a bipartite graph to the counting of perfect
matchings of an extended  bipartite graph. Hence, the monomer-dimer
constants in any dimensions can be computed by permanents of
matrices. Permanent of matrix is studied for a very long time
 \cite{Min78,LP86}.
After Valiant proves that evaluating the permanent of a 0-1 matrix
is a $\#P$-complete
problem \cite{Val79}, many randomized approximate algorithms are developed \cite{KKLLL93,CRS03,JSV04}. 
They can give  reasonable estimations for permanent within a
acceptable computer time.

We consider cubic lattice with
periodic condition, and concentrate on two and three dimensional
lattices in the computation. The algorithms are applicable to other
dimensions and domains other than rectangle. For simplicity of
notation, we assume that $m_1=m_2=\cdots=m_d$. But this is not
essential for the algorithms.

In the next section, the formulation of the monomer-dimer
configuration space in matrix permanent is presented. Computational
methods are discussed in section {III}. The sequential importance
sampling algorithms are given to compute matrix permanents. 
In section IV, numerical results are presented which clearly shows
the efficiency of our formulation and the computational methods.
Finally in section V, some discussions and comments are given.

\section{Formulation in Permanent}
Consider each point/{site} in the lattice as a vertex, and an edge
exists if the two vertices are {neighbors} in the lattice. Hence a
graph $G=(V,E)$ is naturally defined. Using the terminology of graph
theory, a monomer-dimer system can be represented as a covering of
the vertices of the graph $G=(V,E)$ by a non-overlapping arrangement
of monomers (molecules covering one vertex) and dimers (molecules
covering a pair of adjacent vertices).

It is convenient to identify monomer-dimer configurations with
matching in the graph $G$. The sites of a cubic lattice can be
divided into two vertex sets $V_1$ and $V_2$. A site and its
neighbor should always belong to different vertex sets. There are
edges between neighbors, and all edges form a edges set $E$. Thus an
undirected bipartite graph $G(V_1\cup V_2, E)$ is constructed. In
terms of the graph theory, a covering of all vertices   with dimers
is a perfect matching of the bipartite graph $G(V_1\cup V_2, E)$;
and a covering  with $k$ dimers is a $k$-matching of  it. Hence the
cardinal number of the configuration space of the monomer-dimer
model equals to the number of all possible matchings of the
bipartite graph $G$.

The partition function of the system is defined as
\begin{equation}
Z(\lambda)\equiv Z_G(\lambda)=\sum\limits^n_{k=0} m_k \lambda^k
\end{equation}
where $m_k=m_k(G)$ is the number of $k-$matching in the graph $G$,
which is equivalent to the number of monomer-dimer configurations
with $k$ dimers. $Z_G(1)$ enumerates all possible matchings in $G$.

Let $G$ be a bipartite graph and $A$ be the adjacent matrix of the
graph $G$. The number of perfect matchings of $G$ is equal to the
{\em permanent} of the matrix $A$, which is defined as
\begin{equation}\label{per}
perm(A)=\sum_{\sigma\in\Pi_n}\prod_{i=1}^{n}a_{i\sigma{(i)}}.
\end{equation}
Here $\Pi_n$ is the symmetric group of degree $n$.

{ A matrix permanent formulation for enumerating $k$-matching of a
bipartite graph is proposed by Friedland and Levy recently
\cite{FL06}.} Their  method can be applied to approximate the
monomer-dimer constant. For any given $k$, method by Friedland and
Levy can compute $m_k$, the number of $k$-matching,  for all $k\in
\{0,1,\cdots,n\}$. Thus $Z_G(1)$, all possible matchings in $G$, can
be given by
\begin{equation}\label{zg}
Z_G(1)=\sum\limits_{k=0}^n m_k .
\end{equation}
Note that the number of matrix permanents computed is $n$, and $n$
would not be a small number. Here in the following we propose a new
formulation in matrix permanent. The number of all possible
matchings, that is $Z_G(1)$ in (\ref{zg}), can be approximated
directly.

Let $A$ be the adjacent matrix of a bipartite graph $G$. Thus $A$ is
a 0-1 matrix. We use $G(A)$ denote the bipartite graph with adjacent
matrix $A$. The vertex set of $G(A)$ is denoted as $V=V_1 \cup V_2$
and the edge set is $E$. An auxiliary graph is constructed based on
graph $G(A)$ as follows. Vertex sets $V_1^{'}$ and $V_2{'}$ are
added to $V_1$ and $V_2$ respectively. The cardinal numbers of the
new sets $V_1^{'}$ and $V_2{'}$ are both $n$. There are $n$ edges
between $V_1$ and $V_2^{'}$ and each vertex in $V_1$ is adjacent to
a different vertex in $V_2^{'}$. The vertexes of $V_1^{'}$ are
adjacent to every vertexes of $V_2$ and $V_2^{'}$. Let
\begin{equation}\label{mat-B}
B= \left(
\begin{array}{cc}
A & I_{n\times n}\\
1_{n\times n} & 1_{n\times n}
\end{array}\right),
\end{equation}
where $1_{n\times n}$ is the $n\times n$ matrix whose  entries are
all equal to $1$; and $I_{n\times n}$ is the identity matrix of
order $n$. It is obvious that $B$ is a 0-1 matrix, and it is the
adjacent matrix of the auxiliary graph.

Let $AM(A)$ denote the number of  all possible matchings of the
graph $G(A)$. Note that $perm(B)$ gives the number of perfect
matchings of $G(B)$. In a perfect matching of $G(B)$, each vertex
$V_1$ is assigned to be adjacent to a vertex in $V_2\bigcup
V_2^{'}$. The number of all the possible assignment between $V_1$
and $V_2\bigcup V_2^{'}$ equals to $AM(A)$. If the adjacent edges
between the set $V_1$ and set $V_2\bigcup V_2^{'}$ are chosen, there
are $n!$ possibilities for choosing the adjacent edges between
$V_1^{'}$ and $V_2\bigcup V_2^{'}$. So we have $AM(A)\cdot
n!=perm(B),$ that is
\begin{equation}\label{transform}
 AM(A) = \frac{1}{n!}\ perm\left(
\begin{array}{cc}
A & I_{n\times n}\\
1_{n\times n} & 1_{n\times n}
\end{array}\right). \end{equation}
\noindent Denote
\begin{equation*}
f(\lambda)= \frac{1}{n!}\ perm \left(
\begin{array}{cc}
A & \lambda \cdot I_{n\times n}\\
1_{n\times n} & 1_{n\times n}
\end{array}\right) =\sum_{k=0}^{n} f_k \lambda^k.
\end{equation*}
\noindent Let $m_k=m_k(G(A))$ be the number of $k-$matching in the
graph $G(A)$. It is easy to verify that,
\begin{equation}
m_k=f_{n-k}.
\end{equation}

Thus we can get the following permanent formulation of the partition
function of monomer-dimer system.
\begin{equation}
Z(\lambda)\equiv Z_G(\lambda)=\sum\limits^n_{k=0} f_{n-k} \lambda^k
.
\end{equation}
Hence the partition function of the monomer-dimer system is
formulated as matrix permanent. It is important to notice that the
matrix $B$ is very special in structure, which will be explored in
the following numerical algorithms.

\section{Computational Methods through Permanent}

Matrix permanent is a long-studied mathematical problem in its own
right \cite{Min78,LP86}.
A bridge between the computation
of permanent and monomer-dimer constant is established via the
relationship \eqref{transform}. Thus the monomer-dimer constant can
be computed by taking the advantage of the efficient algorithms in
matrix permanent.

The definition of the permanent $perm(A)$ looks similar to that of
the determinant $det(A)$. However it is much harder to be computed.
Valiant \cite{Val79} proves that computing a permanent is a
$\#P$-complete problem in counting, even for 0-1 matrices. Hence
approximate algorithms, which can give a reasonable estimation for
$perm(A)$ within acceptable computer time, attract much attentions
recently.

Practical approximate methods for matrix permanents are Monte Carlo
algorithms. One way to do so is to relate matrix permanents to
matrix determinants by randomizing the elements of matrices
\cite{KKLLL93,CRS03}.
The Markov chain Monte Carlo approach can give a fully
polynomial randomized approximation scheme for the permanent of any
arbitrary nonnegative matrix. This is obtained by M.Jerrum,
A.Sinclair and E.Vigoda \cite{JSV04}.
Beichl, O'Leary and Sullivan \cite{BOS01} compute the number of
$k$-matching of monomer-dimer model using Markov chain Monte Carlo
method. They improve the KRS method \cite{KRS96}.

The Monte Carlo methods with sequential importance sampling, which
are a kind of efficient algorithms for approximating permanent, seem
to be promising for the monomer-dimer problem
\cite{Liu01,Ras94,BS99}. Beichl and Sullivan give the best known
numerical result for 3-D dimer constant by using the techniques
\cite{BS99}. The framework of sequential importance sampling for the
permanent of a 0-1 matrix $A$ is as follows.

\

\noindent {\bf Algorithm SIS-P}

\noindent {\bf Step 1.}\ Choose a
nonzero element from the first row of the matrix $A$ with some
probability $p_1$. Suppose the column index of this element be
$k_1$. Set all the other entries in the first row and the $k_1$th
column to 0's;

\noindent {\bf Step 2.}\ Proceed to the next row, applying the
 same sampling strategy as step 1 recursively. Hence the values $p_2, \cdots,
 p_n$ can be obtained;

\noindent {\bf Step 3.}\ Compute
$X=\frac{1}{p_1}\cdot\frac{1}{p_2}\cdot \cdots \frac{1}{p_n}.$

\

The output $X$ of Algorithm SIS-P is a random variable. It is an
unbiased estimator to the permanent of 0-1 matrix $A$. Different
strategies of choosing the probability distributions would lead to
different sequential importance sampling algorithms.

Now {we} apply Algorithm SIS-P to compute the permanent of the
matrix $B$ in (\ref{mat-B}). The matrix structure is so special that
all the elements in the $(n+1)$th to $(2n)$th rows of $B$  are $1$.
Hence {\bf only} the first $n$ rows of $B$ {are needed to be
considered}. Assume that one sampling gets probability values $p_1,
p_2, \cdots, p_n$. The sampling value should be assigned as
$$\frac{1}{p_1}\cdot\frac{1}{p_2}\cdot \ldots \frac{1}{p_n}\cdot
n!.$$

\noindent If $N$ samples are obtained by Algorithm SIS-P, the number
of all matchings can be approximated by
$$AM(A)= \frac{perm(B)} {n!} \approx
\sum_{j=1}^{N}\frac{1}{p_1^{(j)}}\cdot\frac{1}{p_2^{(j)}}\cdot
\ldots \frac{1}{p_n^{(j)}}.$$

\begin{figure}
\caption{\label{fig:com} The lattice is a $(4,4)$ lattice and thus
the adjacent matrix is $16\times 16$. The x-axis denotes the number
of samplings and the y-axis denotes the error of the approximate
cardinal number of configuration space.}
\begin{center}
\includegraphics[scale=0.3]{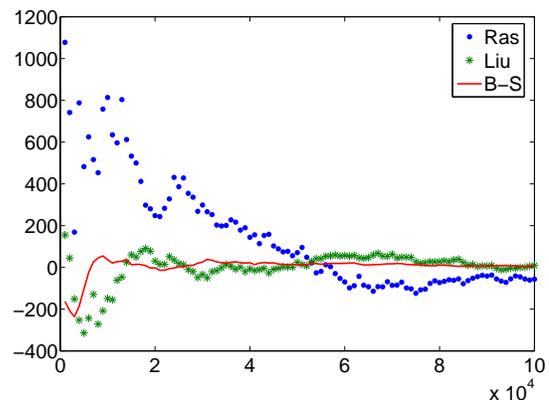}
\end{center}
\end{figure}

Three different importance sampling methods Ras by \cite{Ras94}, Liu
by \cite{Liu01}, and B-S by \cite{BS99} are used respectively  to
compute the number of the cardinal number of the configuration
space.
The results are given in Table
\ref{tab:tablecom}. The convergence rates of the three algorithms
for $m=4$  are also shown in FIG.~\ref{fig:com}. Simple examples
show that both Liu and B-S give good results, and Liu runs faster in
the computation of monomer-dimer constant.

\begin{table}
\caption{\label{tab:tablecom} Comparison of three SIS algorithms for
small 2-dimensional lattice. $m$ denotes the size of $(m,m)$
lattice. Every algorithm sample 10,000 samples. Value denotes the
approximate cardinal number of configuration space of the lattice,
and computer times are given in seconds. }
\begin{ruledtabular}
\begin{tabular}{c|cc|cc|cc|c}
& \multicolumn{2}{c|}{Ras} & \multicolumn{2}{c|}{Liu} &
\multicolumn{2}{c|}{B-S}
\\
\hline m & value & time(s)  & value & time(s) & value & time(s)&
exact value
\\
\hline 2 &  6.9999 & 1.72 & 7.0000 &1.99&  7.0017 &4.164 & 7
\\
\hline 4 &  41280  & 4.14 & 41225 & 5.55 & 41985& 19.47 & 41025
\end{tabular}
\end{ruledtabular}
\end{table}

According to the law of large numbers, the mean value of these
samples gives an approximation to the permanent. But in fact, the
number of samples in our computation is not really ``large''. More
precisely, a typical sample number in our computation would be $100,
000$, while the cardinal number of the sample space could be, for
example, $10^{115}$ (the two dimensional monomer-dimer model with
$m=20$).




Notice that the probability distribution of the random variable
$Y=\log X$ looks similar to the normal distribution. If the
probability distribution of $Y$ is 
normal with $N(\mu,
\sigma)$, then the expectation of $X$ would be
\[E(X)=E(e^Y)=e^{\mu+\frac{\sigma^2}2}.\]
Other than computing the sample mean of $X$ directly, we can
estimate the sample mean $\bar{\mu}$ and sample standard deviation
$\bar{\sigma}$ of the random variable $Y$ first.

\section{Experimental Results for Periodic Lattices}

The algorithm SIS is
used to approximate permanents, which gives approximation to the
monomer-dimer constants. {The algorithms are programmed in Matlab
7.0 and all computations in this paper run on Dell PC with CPU 2.8G
Hz.}

\subsection{Experiments on two dimensional lattices}

Computational results for 2-dimensional monomer-dimer problems with
periodic boundary conditions are presented in TABLE
\ref{tab:table1}.


\begin{table}[!h]
\caption{\label{tab:table1} $m$ denotes the size of the planar
$(m,m)$ lattice. Every time, we sample $100,000$ samples and compute
the approximate result of $log Z(m,2)/m^2$. We do this several
times. SIS gives the median value of the approximate values;  Time
denotes the time in second for one sampling. A-PRE is the value
given in \cite{BOS01}}
\begin{ruledtabular}
\begin{tabular}{cccc}
m & SIS & Time(sec) & A-PRE\\
\hline
 4  & 0.663866 &   0.0012&   0.6611  \\
 6  & 0.662851 &   0.0019&   0.6629\\
 8  & 0.662897 &   0.0028&   0.6611\\
 10 & 0.662951 &   0.0038&   0.6663\\
 12 & 0.662990 &   0.0055&   0.6646\\
 14 & 0.662852 &   0.0072&   0.6638\\
 16 & 0.662644 &   0.0100& ---\\
 18 & 0.663390 &   0.0138& ---\\
 20 & 0.662960 &   0.0181& ---\\
 22 & 0.663031 &   0.0237& ---\\
 24 & 0.662893 &   0.0307& ---\\
 26 & 0.663754 &   0.0398& ---\\
 28 & 0.663013 &   0.0507& ---\\
 30 & 0.663062 &   0.0710& ---\\
 32 & 0.662587 &   0.0769& ---

\end{tabular}
\end{ruledtabular}
\end{table}



\begin{figure}[!h]
\caption{\label{fig:fit} The relations between the running time of
A-PRE method and SIS method with the lattice size $m$ are shown. The
times of SIS method are costs of $100,000$ samples, those of A-PRE
method are taken from \cite{BOS01}.}
\begin{center}
\includegraphics[scale=0.6]{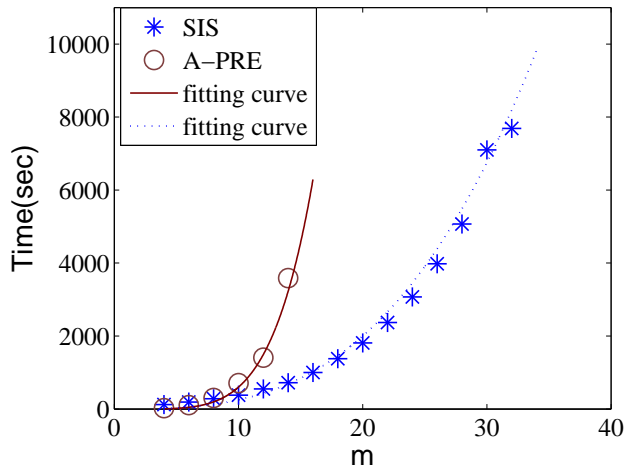}
\end{center}
\end{figure}

Let compare with the results and the algorithm A-PRE, a Markov Chain
Monte Carlo method used by Beichl, O'Leary, and
Sullivan\cite{BOS01}.  Though computers used here are different, one
can still tell the trends in the running times. The curve fitting
results for algorithms SIS and A-PRE are shown in FIG \ref{fig:fit}.
It is clear that the running times for both SIS and A-PRE grow
polynomially with respect to $m$. The time complexity of SIS, the
method developed in this paper, is about $O(m^{3})$ for
2-dimensional lattice, while the A-PRE, the MCMC method by
\cite{BOS01}, is about $O(m^{5})$. Hence it is  easy to tell that
the algorithm SIS is faster distinctly. This suggests that the
method SIS can be applied to large monomer-dimer problems.

In order to fit the limit of $log Z(m,2)/m^2$ as $m$ goes to
infinity, we apply regression to the computed mean values. The
regression function is the same as \cite{BS99}
\begin{equation}\label{reg-2}
y=\frac{p_1}{x^2}+p_2,
\end{equation}
where $x$ denotes the lattice size $m$, $y$ denotes the $h_2(m)$ and
$p_2$ is the monomer-dimer constant. The monomer-dimer constant of
2-dimensional problem with periodic boundary can be obtained from
the regression
$$h_2=0.6627\pm 0.0002\ \ with\ \ 95\%\ \ confidence.$$
\noindent The approximate results of the monomer-dimer constant
coincides with the   value  $h_2=0.662798972834$ by \cite{Kong06}
very well.

\subsection{Experiments on three dimensional lattices}

For 3-dimensional 
problem with periodic condition, computational results  are shown in
TABLE \ref{tab:table3}. The time complexity for algorithm SIS in 3
dimensional problems is about $O(m^{6})$.

\begin{table}[!h]
\caption{\label{tab:table3} $m$ denotes the size of the cubic
$(m,m,m)$ lattice. Every time, we sample $100,000$ samples and
compute the approximate result of $log Z(m,3)/m^3$. We do this
several times. SIS gives the median value of the approximate values;
Time denotes the time in second for one sampling. A-PRE is the value
given in \cite{BOS01}  }
\begin{ruledtabular}
\begin{tabular}{cccccc}
m & SIS &  Time(sec) & A-PRE\\
\hline
4&   0.787359&    0.0039&     0.7844\\
6&   0.786661&    0.0082&     0.7847\\
8&   0.785821&    0.0345&     0.7870\\
10&  0.787093&    0.0919&   ---\\
12&  0.785054&    0.2483&   ---\\
14&  0.783476&    0.6693&   ---
\end{tabular}
\end{ruledtabular}
\end{table}


To fit the limit of $log Z(m,3)/m^3$ as $m$ goes to infinity, we
apply regression again. The function we use is
\begin{equation}\label{reg-3}
y=\frac{p_1}{x}+p_2,
\end{equation}
where $x$ denotes the lattice size $m$, $y$ denotes the $h_3(m)$ and
$p_2$ is the monomer-dimer constant. The result is
$$h_3=0.7847\pm 0.0014\ \ with\ \ 95\%\ \ confidence.$$
This agrees well with the best known bound $0.7653 \leq h_3 \leq
0.7862$ \cite{FP05}.

\section{Discussions and Comments}

The construction of the auxiliary bipartite graph is the key step in
our formulation. Hence the permanent of the matrix $B$ in
(\ref{mat-B}) gives the total number of matchings in the original
bipartite graph $G(A)$. The size of the matrix $B$ doubles that of
$A$. However since the special structure of the matrix $B$ can be
explored in the algorithm,  the computational cost does not really
increase. 

The Monte Carlo method we used in this paper is based on the
sequential importance sampling. Each time one samples a term from
the large sum defined by (\ref{per}), and only nonzero terms are
valuable in the computation. The formulation and computational
methods 
that we propose in this paper never meet any zero term. This fact is
not obvious but crucial for the efficiency of the algorithm. A
rigorous mathematical proof  will be presented elsewhere.

The regression function (\ref{reg-2}) for two dimensional  is
discussed and used by many authors, for example \cite{BS99}. It is
based on asymptotic analysis. However (\ref{reg-3}) for three
dimensional is just a result of statistical experiments. We are
unable to give it any physical reasoning.

The basic contribution of this paper is the  formulation and
computational methods for approximating the number of all matchings
in bipartite graphs. In this way, larger monomer-dimer systems can
be studied.

\

\begin{acknowledgments}
We wish to acknowledge the support of by National Science Foundation
of China {10501030}.
\end{acknowledgments}


\parskip 0pt
{\baselineskip=12pt \small{
}

\end{document}